\documentclass{JHEP3} 
\pdfoutput=1
\usepackage{epsfig,amssymb, graphicx,bm,color, subfigure}
\let\normalcolor\relax   

\usepackage{graphicx}
\usepackage{url}
\usepackage{amssymb,amsmath}
\usepackage{amsbsy}

\def\Msun{M_\odot}
\def\hmpcinv{h\,{\rm Mpc}^{-1}}
\def\hmpc{\,h^{-1}\,{\rm Mpc}}

\newcommand{\fnl}{f_{\rm NL}}

\newcommand{\perm}{{\rm perm.}}

\newcommand\ba{\begin{eqnarray}}
\newcommand\ea{\end{eqnarray}}

\newcommand{\eqn}[2]{
\begin{equation}\label{#1} 
#2 
\end{equation}
}

\title{Scale-Dependent Non-Gaussianity as a Generalization of the Local Model}

\author{Adam Becker, Dragan Huterer, Kenji Kadota\\
{\small Department of Physics and Michigan Center for Theoretical Physics\\
 University of Michigan, 450 Church Street, Ann Arbor, MI 48109}}

\abstract{We generalize the local model of primordial non-Gaussianity by
  promoting the parameter $\fnl$ to a general scale-dependent function
  $\fnl(k)$. We calculate the resulting bispectrum and the effect on the bias
  of dark matter halos, and thus the extent to which $\fnl(k)$ can be measured
  from the large-scale structure observations.  By calculating the principal
  components of $\fnl(k)$, we identify scales where this form of
  non-Gaussianity is best constrained and estimate the overlap with
  previously studied local and equilateral non-Gaussian models. 
}


\begin{document}

\section{Introduction}\label{sec:intro}

Primordial non-Gaussianity provides cosmology one of the precious few
connections between primordial physics and the present-day universe. Standard
inflationary theory, with a single slowly rolling scalar field, predicts that
the spatial distribution of structures in the universe today is very nearly
Gaussian random (e.g.\
\cite{maldacena,Acquaviva:2002ud,Creminelli:2003iq,Lyth_Rodriguez,Seery_Lidsey};
for excellent recent reviews, see \cite{Chen_AA,Komatsu_CQG}).  Departures
from Gaussianity, barring contamination from systematic errors or late-time
non-Gaussianity due to secondary processes, would be a violation of this
standard inflationary assumption.  Constraining or detecting primordial
non-Gaussianity is therefore an important basic test of the standard
cosmological model.

Most of the study of non-Gaussianity in the literature to date has been
carried out assuming the magnitude of departure from Gaussianity is
scale-independent
(e.g.\ \cite{Komatsu_Spergel,Verde_CMBLSS,Scoccimarro:2003wn}).  However, the
assumption that $\fnl$ is constant for a wide range of scales could be an
over-simplification, since the primordial cosmic perturbations were presumably
produced from the time-dependent dynamics in the early universe. In
particular, single-field inflationary models with interactions, along with
most multi-field models, generically produce scale-dependent
non-Gaussianity. It is therefore not surprising that scale-dependence of
non-Gaussianity has been discussed in the community in recent years
\cite{Salopek,Falk_Ran_Sre,Luo_Schramm,Gangui_etal,Wang_Kam,Bartolo:2004if,see2, Chen2005,Liguori2006,Chen2006,LoVerde,Chen2008,Sefusatti2009,Kumar2010,Byrnes2010,bryb,wandsb, Riotto2010,Huang2010}. Notably,
the parameterization of the scale-dependent non-Gaussianity in our analysis is
applicable to the curvaton \cite{mol,lin5,enq5,ly5,moro} and the modulated
reheating scenarios \cite{lev,mati}, which are of great interest for their
potentially observable scale-dependent non-Gaussianity\footnote{For instance,
  when the observed perturbations originate from the single
  curvaton field, the ``running'' (with scale) of the non-Gaussianity
  parameter is proportional to the third derivative of the curvaton potential,
  $V'''$ \cite{chris5,chris6,huang}. Given that this third derivative is not
  tightly constrained from the observed power spectrum, it can potentially
  lead to observable {\it and} scale-dependent non-Gaussianity. Therefore,
  constraints on the running of non-Gaussianity can be a powerful probe of the
  origin of the primordial curvature perturbations.}.


Motivated by such inflationary models that predict detectable scale-dependent
non-Gaussianity, as well as a desire to have an easily usable basis for studying
those models, we present a novel scale-dependent ansatz for primordial
non-Gaussianity: we promote the parameter $\fnl$ to a free
function of wavenumber $\fnl(k)$. We define our model (Sec.~\ref{sec:NG}),
predict clustering bias of dark matter halos in our model
(Sec.~\ref{sec:NG_bias}), obtain an upper bound on the accuracy with which
these new parameters could be measured with a future large-scale structure
survey (Sec.~\ref{sec:forecasts}), and compare our model with other
parameterizations of non-Gaussianity in the literature (Sec.~\ref{sec:PC}).

\section{Scale dependent non-Gaussianity}
\label{sec:NG}



The most commonly discussed model of non-Gaussianity, often referred to as the
local model, is defined via \cite{Komatsu_Spergel}
\eqn{eq:localNG}{
\Phi(x)=\phi_G(x)+\fnl(\phi_G(x)^2-\langle \phi_G (x)^2 \rangle ).
}
Here, $\Phi$ denotes the primordial curvature perturbations (Bardeen's
gauge-invariant potential), $\phi_G(x)$ is a Gaussian random field, and the
constant $\fnl$ is the non-Gaussianity parameter. The local model has been
much studied, in part because it is the first two terms of the most general
local form of non-Gaussianity \cite{Babich_shape}.

In Fourier space, Eq.~(\ref{eq:localNG}) becomes
\begin{equation}
\Phi(k)=\phi_G(k)+\fnl\int \frac{d^3 k'}{(2 \pi)^3}\phi_G(k')\phi_G(k-k').
\label{eq:localNG_kspace}
\end{equation}
(Hereafter, we omit the subscript $G$ on the Gaussian distribution when it is
clear from context.) In this paper, we study a model that generalizes Eq.~(\ref{eq:localNG_kspace}) -- we allow $\fnl$ to vary with $k$ as well, while
assuming isotropy and homogeneity (so $\fnl(${\boldmath $k$}$)=\fnl(k)$). The
gravitational potential in the new model is defined via
\begin{equation}
\Phi(k)=\phi(k)+\fnl(k)\int \frac{d^3 k'}{(2 \pi)^3}\phi(k')\phi(k-k').
\label{eq:fnlk_kspace}
\end{equation}
As mentioned above, this form of non-Gaussianity is expected in curvaton or
modulated reheating scenarios (see e.g.\ Ref.~\cite{chris5}, where this form
explicitly appears in the study of these models).

Note that this new ansatz is {\it not} local, which is clear when we transform
back into real space:
\begin{equation}
\Phi(x)=\phi+\fnl(x)*(\phi(x)^2-\langle \phi (x)^2 \rangle ),
\label{eq:fnlk_realspace}
\end{equation}
where $*$ represents convolution and $x$ denotes a three-dimensional spatial
coordinate. These primordial
perturbations $\Phi(k)$ are related to the present-time (z=0) smoothed linear
overdensity $\delta_R$ by the Poisson equation:
\begin{equation}
\delta_R(k)=\frac{2}{3} 
\frac{k^2 T(k)}{H_0^2 \Omega_{m}} \tilde{W}_R(k) \Phi(k) 
\equiv \mathcal{M}_R(k) \Phi(k);
\label{eq:overdensity}
\end{equation}
where $T(k)$ is the matter transfer function, $H_0$ is the Hubble constant,
$\Omega_m$ is the matter density relative to critical today, and
$\tilde{W}_R(k)$ is the Fourier transform of the top-hat filter with radius
$R$. The smoothing spatial scale $R$ is related to the smoothing mass scale
$M$ via
\begin{equation}
M = {4 \over 3} \pi R^3 \rho_{m, 0},
\label{eq:mass_scale}
\end{equation}
where $ \rho_{m, 0}$ is the matter energy density today.  The choice of mass
scale is discussed further in section \ref{sec:constraints}.

The bispectrum in our generalized model becomes
\begin{equation}
B_{\phi}(k_1, k_2, k_3) = 2[\fnl(k_1) P_\phi(k_2)P_\phi(k_3) + \perm],
\label{eq:fnlk_bispec}
\end{equation}
where $P_\phi$ is the power spectrum of potential fluctuations.
This reduces to the familiar expression $B(k_1, k_2, k_3) = 2 \fnl (P_\phi(k_1)P_\phi(k_2) + \perm)$
when $\fnl$ is a constant.


Notice the difference between our ansatz for the scale-dependent $\fnl(k)$ (which has the corresponding bispectrum Eq.~(\ref{eq:fnlk_bispec})) and
the particular form of scale-dependent non-Gaussianity, discussed elsewhere in the
literature, which is defined as $\fnl(k_1,k_2,k_3)\equiv B_\phi(k_1, k_2, k_3)/[2
  P_\phi(k_1)P_\phi(k_2) + \perm]$ (\cite{Byrnes2010,wandsb,bryb}). The two forms are inequivalent, and either
form can be borne out in realistic inflationary models; however, given that
our form lives in a lower-dimensional $k$-space, it is easier to simulate it
numerically \cite{Shandera2010} or treat it with the Fisher matrix analysis, as we
do in this paper.

\section{Non-Gaussianity and Bias}
\label{sec:NG_bias}

\subsection{The effect of a non-vanishing bispectrum on bias}

Dalal et al.\ \cite{Dalal} found, analytically and numerically, that the bias of dark
matter halos acquires strong scale dependence if $\fnl \neq 0$:
\begin{equation}
b(k)=b_0 + \fnl(b_0-1)\delta_c\, \frac{3\Omega_mH_0^2}{a\,g(a) T(k)c^2 k^2}.
\label{eq:bias}
\end{equation}
Here, $b_0$ is the usual Gaussian bias (on large scales, where it is
constant), $\delta_c\approx 1.686$ is the collapse threshold, $a$ is the scale
factor, $\Omega_m$ is the matter density relative to the critical density, $H_0$ is the
Hubble constant, $k$ is the wavenumber, $T(k)$ is the transfer function, and
$g(a)$ is the growth suppression factor\footnote{The usual linear growth
  $D(a)$, normalized to be equal to $a$ in the matter-dominated epoch, is
  related to the suppression factor $g(a)$ via $D(a)=ag(a)$, where $g(a)$ is
  normalized to be equal to unity deep in the matter-dominated epoch.}. This
result has been confirmed by other researchers using a variety of methods,
including the peak-background split
\cite{Afshordi_Tolley,MV,Slosar_etal,Schmidt_Kam}, perturbation theory
\cite{McDonald,Taruya08,GP}, and numerical (N-body) simulations
\cite{Grossi,Desjacques_Seljak_Iliev,PPH}.  Astrophysical measurements of the
scale dependence of the large-scale bias, using galaxy and quasar clustering
as well as the cross-correlation between the galaxy density and CMB
anisotropy, have recently been used to impose constraints on $\fnl$ already
comparable to those from the cosmic microwave background (CMB) anisotropy
\cite{Slosar_etal, Afshordi_Tolley}, giving $\fnl=28\pm 23$ ($1\sigma$), with
some dependence on the assumptions made in the analysis \cite{Slosar_etal}.
In the future, constraints on $\fnl$ are expected to be on the order of a few
\cite{Dalal,Carbone,Sartoris,Cunha_NG}.  The sensitivity of the large-scale
bias to other models of primordial non-Gaussianity has not yet been
investigated much (though see analyses in e.g.~\cite{Desjacques_gnl,MV09}).

Following the MLB formula \cite{Grinstein:1986en,MLB1986}, one can express
the two point correlation function of dark matter halos,
$\xi_h(\boldsymbol{x}_1,{\boldsymbol x}_2 )$, in terms of certain
configurations of the correlation functions of the underlying density field,
$\xi_R^{(N)}$. In the high-threshold limit ($\nu \gg 1$), this becomes:
\begin{eqnarray}
\xi_h(\boldsymbol{x}_1,{\boldsymbol x}_2 ) &=&\xi_h(x_{12}) \nonumber\\&=&
-1+\exp
\left(
\sum_{N=2}^{\infty} \sum_{j=1}^{N-1} \frac{\nu^N}{\sigma_R^N} \frac{1}{j! (N-j)!}\xi_R^{(N)}
\left[
\begin{array}{cc} 
\boldsymbol{x}_1,...,\boldsymbol{x}_1,& \boldsymbol{x}_2,...,\boldsymbol{x}_2 \\
j~ {\rm times}&(N-j)~{\rm times}
\end{array}
\right]
\right);
\label{eq:Grinstein-Wise}
\end{eqnarray}
where $x_{ij}=|\boldsymbol{x}_i-\boldsymbol{x}_j|,\nu=\delta_c/\sigma_R$
represents the peak height, and ${\xi_R}^{(n)}(r)$ is the $n$-point
correlation function of the underlying matter density smoothed with a top-hat
filter of radius $R$. Keeping the terms up to the three-point correlation 
function, which would be reasonable for the observationally allowed range of 
$\fnl$, the expansion series gives us the halo correlation function in terms of the field correlation functions:
\begin{equation}
\xi_h(x_{12})= \frac{\nu^2}{\sigma_R^2}
   {\xi_R^{(2)}}(\boldsymbol{x}_1,\boldsymbol{x}_2)+ \frac{\nu^3}{\sigma_R^3}
   {\xi_R^{(3)}}(\boldsymbol{x}_1,\boldsymbol{x}_1,\boldsymbol{x}_2).
\label{eq:ksi_MV}
\end{equation}

The  Fourier transform of the real-space correlation function -- the power
spectrum -- is given, to the same expansion order as Eq.~(\ref{eq:ksi_MV}), by
\begin{equation}
\label{ph1}
P_h(k)= \frac{\nu^2}{\sigma_R^2}  
P_R(k)+ \frac{\nu^3}{\sigma_R^3}  
\int \frac{d^3 q}{(2 \pi)^3}
B_R(k,q,|\boldsymbol{k}-\boldsymbol{q}|)+\ldots 
\end{equation}
The first term on the right-hand side includes the familiar (Gaussian) bias
$b={\nu}/{\sigma_R}$ (in the high-peak limit for which the MLB formula is valid) 
for the Gaussian fluctuations.  The effects of non-Gaussianity on the
galaxy bias are represented by the second term, including the bispectrum
$B_R$, which vanishes for the Gaussian fluctuations.

\subsection{From the bispectrum to bias}

If we denote the full bias of dark matter halos by $b+\Delta b$, where
$b$ represents the bias for the Gaussian fluctuations and $\Delta b$ is the
non-Gaussian correction, then 
\begin{equation}
\frac{P_h}{P_R}=b^2\left(1+\frac{\Delta b}{b}\right)^2,
\label{eq:bL}
\end{equation}
where $P_h$ and $P_R$ are the power spectra of halos and dark matter,
respectively.  The non-Gaussian correction to the linear peak bias to the
leading order becomes
\begin{equation}
\frac{\Delta b}{b} (k) =\frac{\nu}{\sigma_R} \,\frac{1}{2 P_R(k)} \int \frac{d^3
  q}{(2 \pi)^3} B_R(k,q,|\boldsymbol{k}-\boldsymbol{q}|),
\end{equation}
where $B_R$ is the matter bispectrum on scale $R$.
Hence, the non-Gaussian correction 
$\Delta b(k)$ can be expressed in terms of the primordial potential fluctuations as (\cite{MV}):
\begin{equation}
{\Delta b\over b} (k) =
{\delta_c\over D(z)}\, {1\over 8\pi^2 \sigma_R^2 \mathcal{M}_R(k)}
\int_0^{\infty} dk_1 k_1^2 \mathcal{M}_R(k_1)\, \int ^{1}_{-1}d\mu \mathcal{M}_R(k_2) 
{B_{\phi}(k_1, k_2, k)\over P_{\phi}(k)}.
\label{eq:MV}
\end{equation}
We perform the integration over all triangles. The triangles' sides are
$k_1$, $k_2$, and $k$; the cosine of the angle opposite $k_2$ is $\mu$, so $k_2^2 = k_1^2 +
k^2 + 2k_1k\mu$.  $\mathcal{M}_R (k)$ is the same function defined in
Eq.~(\ref{eq:overdensity}), and the time dependence of the critical
threshold for collapse is given as $\delta_c(z) = \delta_c/D(z)$, with
$\delta_c=1.686$.

\subsubsection {Constant $\fnl$}

Eq.~(\ref{eq:MV}) leads to the famous scale-dependent
bias formula in the case of a constant $\fnl$. For this model, the bispectrum is
\begin{equation}
B_\phi(k_1, k_2, k_3) = 2\fnl\, [P_\phi(k_1)P_\phi(k_2) + \perm].
\label{eq:local_bispec}
\end{equation}
Through Eq.~(\ref{eq:MV}), this leads to the result
\begin{eqnarray}
{\Delta b\over b} (k) &=& \nonumber
{\delta_c\over D(z)}\, {2\fnl\over 8\pi^2 \sigma_R^2 \mathcal{M}_R(k)}
\int dk_1 k_1^2 \mathcal{M}_R(k_1)P_\phi(k_1) 
\int d\mu \mathcal{M}_R(k_2) \left [ {P_\phi(k_2)\over P_\phi(k)} + 2\right ]\\[0.3cm]
 &\equiv & {2\fnl \delta_c\over D(z)}\,{\mathcal{F}(k)\over  \mathcal{M}_R(k)},
\label{eq:dboverb_fnlconst}
\end{eqnarray}
where
\begin{equation}
\mathcal{F}(k)\equiv {1\over 8\pi^2 \sigma_R^2}
\int dk_1 k_1^2 \mathcal{M}_R(k_1)P_\phi(k_1)
\int d\mu \mathcal{M}_R(k_2) \left [ {P_\phi(k_2)\over P_\phi(k)} + 2\right ].
\end{equation}
Note that there is a factor of $2$ in Eq.~(\ref{eq:dboverb_fnlconst}) because we can exchange the order of integration of terms corresponding to $k_1$
and $k_2$.

Finally, we rewrite Eq.~(\ref{eq:dboverb_fnlconst}) by defining
\begin{eqnarray}
\mathcal{F}_1(k)&\equiv& {1\over 8\pi^2 \sigma_R^2 \mathcal{M}_R(k) P_\phi(k)}
\int dk_1 k_1^2 \mathcal{M}_R(k_1)P_\phi(k_1)
\int d\mu \mathcal{M}_R(k_2) P_\phi(k_2)
\label{eq:F1_dummy}
\\[0.2cm]
\mathcal{F}_2(k)&\equiv& {2\over 8\pi^2 \sigma_R^2 \mathcal{M}_R(k)}
\int dk_1 k_1^2 \mathcal{M}_R(k_1)P_\phi(k_1)
\int d\mu \mathcal{M}_R(k_2).
\label{eq:F2_dummy}
\end{eqnarray}
Then, for constant $\fnl$,
\begin{equation}
{\Delta b\over b} (k) = 
{2\fnl \delta_c\over D(z)}\,\left [\mathcal{F}_1(k) + \mathcal{F}_2(k)\right ],
\end{equation}
and the derivative with respect to $\fnl$ is
\begin{equation}
{\partial\over\partial \fnl}\left [{\Delta b\over b} (k)\right ]
= {2 \delta_c\over D(z)}\,\left [\mathcal{F}_1(k) + \mathcal{F}_2(k)\right ].
\end{equation}

\subsubsection{Scale-dependent  $\fnl$}

Now we repeat the analysis of the previous section, but we allow $\fnl(k)$ to be an
arbitrary function of scale, adopting the ansatz in
Eq.~(\ref{eq:fnlk_kspace}). We still assume homogeneity, so
$\fnl(\vec{k})=\fnl(k)$. The bispectrum is given by
\begin{equation}
B_{\phi}(k_1, k_2, k_3) = 2[\fnl(k_1) P_\phi(k_2)P_\phi(k_3) + \perm].
\end{equation}
Here, the triangle condition always holds, so that (for example) $k_1 =
|\vec{k_2}+\vec{k_3}|$. Following Eq.~(\ref{eq:MV}), we get
\begin{eqnarray}
{\Delta b\over b} (k)& =&
{\delta_c\over D(z)}\, {2\over 8\pi^2 \sigma_R^2 \mathcal{M}_R(k)}
\int dk_1 k_1^2 \mathcal{M}_R(k_1)P_\phi(k_1)
\nonumber \\[0.2cm]
&\times &\int d\mu \mathcal{M}_R(k_2) \left [ \fnl(k){P_\phi(k_2)\over P_\phi(k)} + 2\fnl(k_2)\right ].
\end{eqnarray}

This looks like Eq.~(\ref{eq:dboverb_fnlconst}) -- but this time, $\fnl(k)$ is a function, not a constant. Thus, to find the derivative of $\Delta b/ b (k)$ with respect to the relevant parameters, we must parametrize $\fnl(k)$ in a way that is valid for any general form of $\fnl(k)$. We consider the
piecewise-constant (in wavenumber) parametrization where $\fnl(k)$ is equal to $\fnl^i$ in the $i$th wavenumber bin:
\begin{equation}
\fnl^i\equiv \fnl(k_i).
\label{eq:fnl_piecewise}
\end{equation}
The derivative
of $\Delta b/ b (k)$ with respect to these $\fnl^i$ is:
\begin{eqnarray}
{\partial\over\partial \fnl^j}\left [{\Delta b\over b} (k_i)\right ] &=&
{\delta_c\over D(z)}\, {2\over 8\pi^2 \sigma_R^2
  \mathcal{M}_R(k)}\times\nonumber \\[0.3cm]
&& \left [
\delta_{ij}{1\over P_\phi(k)} 
\int dk_1 k_1^2 \mathcal{M}_R(k_1)P_\phi(k_1)
\int d\mu \mathcal{M}_R(k_2)P_\phi(k_2) +\right .\\[0.2cm]
&&\left . +2\int_{k_2\in k_j} dk_1 k_1^2 \mathcal{M}_R(k_1)P_\phi(k_1)
  \int d\mu \mathcal{M}_R(k_2) 
\right ],\nonumber
\end{eqnarray}
where $\delta_{ij}$ is the Kronecker delta function. Note that the last
integral over $k_2$ only goes over the $j$th wavenumber bin.

This derivative can be rewritten more concisely as
\begin{equation}
{\partial\over\partial \fnl^j}\left [{\Delta b\over b} (k_i)\right ] =
{2\delta_c\over D(z)}\left [\delta_{ij}\mathcal{F}_1(k) + \mathcal{F}^j_2(k)\right ].
\label{eq:fnlk_deriv}
\end{equation}
The functions $\mathcal{F}_1$ and $\mathcal{F}_2$ are defined as in
Eqs.~(\ref{eq:F1_dummy}) and (\ref{eq:F2_dummy}), except that the superscript
in $F^j_2$ indicates that the integral over $k_2$ is to be executed only over
the $j$th wavenumber bin.


\section{Forecasted measurements of the scale-dependent nongaussianity}
\label{sec:forecasts}

\subsection{Fisher Matrix Analysis}
\label{sec:constraints}

With an expression for
$\partial/\partial \fnl^j[(\Delta b/ b) (k_i)]$ in hand
(Eq.~(\ref{eq:fnlk_deriv})), we can calculate the Fisher information matrix for
the parameters $\fnl^j$ that describe the piecewise-constant $\fnl(k)$. The Fisher matrix, in turn, allows us to forecast the extent to which the scale-dependent non-Gaussianity could
be measured in future galaxy surveys. 




We consider measurements of the power spectrum $P_h(k)$ of dark matter halos (galaxies
or clusters, for example) averaged over thin spherical shells in $k$-space. The variance of $P_h(k)\equiv
P_h$ in each shell is \cite{FKP}
\begin{equation}
\label{power_error}
\sigma^2_{P_h} = \frac{2 P_h^2}{V_{\rm shell} \;  V_{\rm survey}} 
\left( \frac{1 + nP_h}{nP_h} \right)^2 = \frac{(2 \pi P_h)^2}{k^2 dk \;  
V_{\rm survey}} \left( \frac{1 + nP_h}{nP_h} \right)^2,
\end{equation}
where $V_{\rm shell} = 4\pi k^2 dk/(2\pi)^3$ is the volume of the shell in
Fourier space (we are ignoring redshift distortion effects for simplicity here).
Therefore, the Fisher matrix for measurements of $P_h(k, z)$ is
\cite{Tegmark97}
\begin{equation}
F_{ij} = \sum_m V_m \int_{k_{\rm min}}^{k_{\rm max}} 
\frac{\partial P_h(k, z_m)}{\partial p_i} \frac{\partial P_h(k, z_m)}{\partial
  p_j} 
\,\frac{1}{\left[ P_h(k, z_m) + \displaystyle\frac{1}{n} \right]^2}\, \frac{k^2 dk}{(2\pi)^2},
\label{PowerFisher}
\end{equation}
where $V_m$ is the comoving
volume of the $m$-th redshift bin, each redshift bin is centered on $z_m$, and we have summed over all redshift bins.
We adopt $k_{\rm min}=10^{-4}\hmpc$, and we choose $k_{\rm
    max}$ as a function of $z$ so that $\sigma(\pi/(2k_{\rm max}), z) = 0.5$
  \cite{SeoEisenstein2003}, which leads to $k_{\rm max}(z= 0)
  \approx 0.1\hmpcinv$.
Finally, $p_{i}$ are the parameters of interest; in our case, these are the
$\fnl^i$. 

We assume a flat universe and a fiducial model of zero non-Gaussianity:
$\fnl(k)=0=\fnl^i$.  We include six cosmological parameters in our Fisher
matrix aside from the $\fnl^i$: Hubble's constant $H_0$; physical dark matter
and baryon densities $\Omega_{\rm cdm} h^2$ and $\Omega_{\rm b} h^2$; equation
of state of dark energy $w$; the log of the scalar amplitude of the matter
power spectrum, $\log A_s$; and the spectral index of the matter power
spectrum, $n_s$ . Fiducial values of these parameters correspond to their
best-fit WMAP7 values \cite{wmap7}. We also added the forecasted cosmological
parameter constraints from the CMB experiment Planck by adding its Fisher
matrix as a prior (W.\ Hu, private communication). Note that the CMB prior
does {\it not} include CMB constraints on non-Gaussianity; the CMB constraints
on $\fnl(k)$ will be separately studied in a future
work. Finally, in addition to the cosmological parameters and
  the $\fnl^i$, we include five Gaussian bias parameters
  in our Fisher matrix -- one $b_0(z)$ for each redshift bin. The fiducial
  values of these parameters are set by the relations $b_0(z =0) = 2.2$, and
  $b_0(z) = b_0(z = 0) / D(z)$.

We already have the derivatives of $b(k)$ with respect to each of the
$\fnl^i$, so the derivative of $P_h(k)$ with respect to the $\fnl^i$ is just
\begin{equation}
\label{BiasDerivative}
\frac{\partial P_{h}(k)}{\partial \fnl^i}
=  2\, \frac{\partial b(k)}{\partial \fnl^i}\, b(k) P_{\rm mat}(k);
\end{equation}
$P_{\rm mat}(k)$ is the $\Lambda$CDM matter power spectrum, easily
obtained from a numerical code such as CAMB. Since we only consider
information from large scales ($k\leq k_{\rm max} \approx 0.1\,\hmpcinv$), we do not
model the small amount of nonlinearity present at the high-$k$ end of these
scales.

%

\begin{figure}[t] 
\begin{center}
	\subfigure[Unmarginalized errors]{\includegraphics[width= 2.9in]{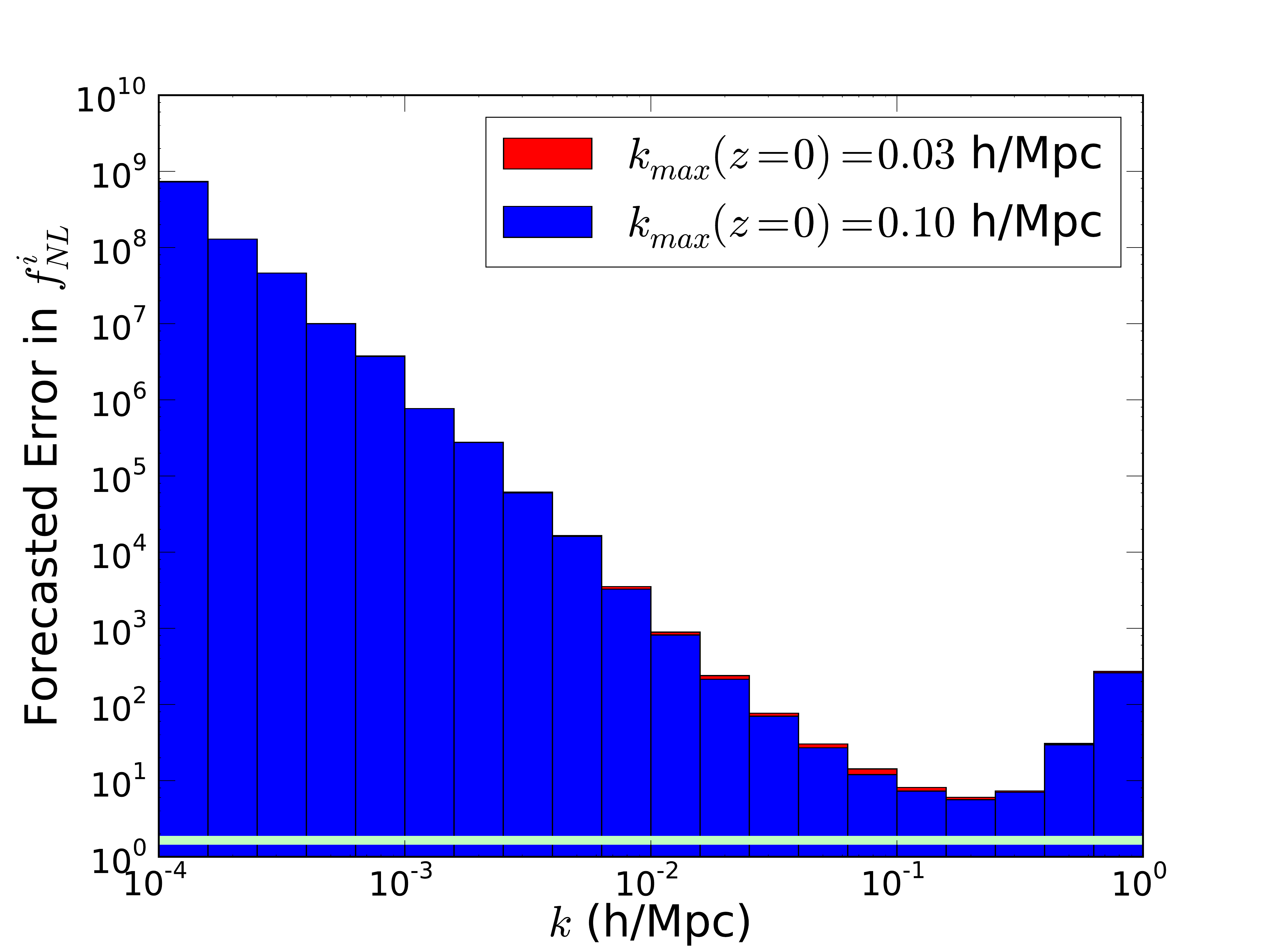}}
	\subfigure[Marginalized errors]{\includegraphics[width= 2.9in]{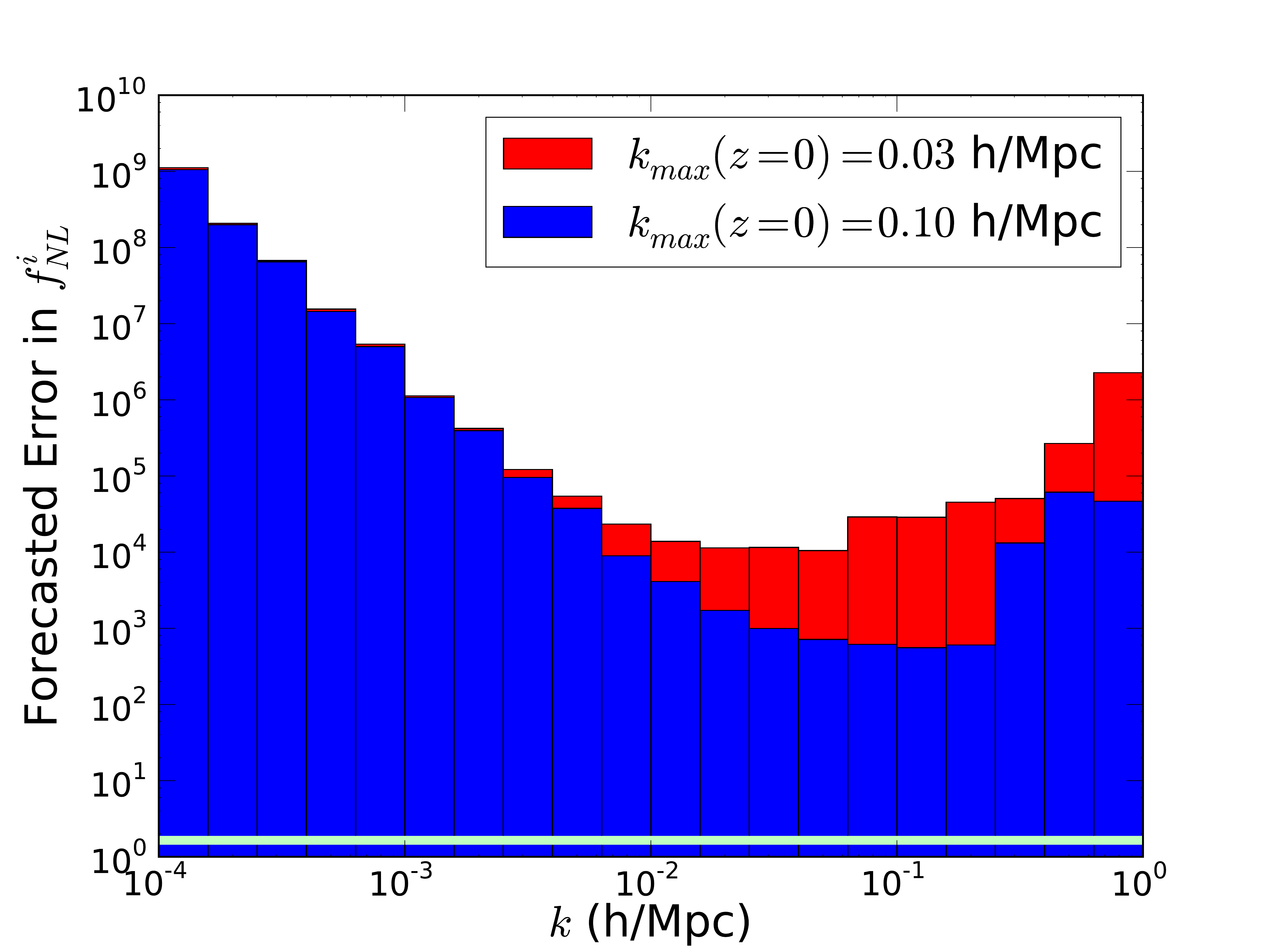}}
        \caption{Estimated unmarginalized (left panel) and marginalized (right
          panel) constraints on piecewise-constant parameters $\fnl^i$
          assuming a future galaxy survey covering one-quarter of the sky out
          to $z = 1$, with average number density of $2\times10^{-4}$
          gal/Mpc$^3$. For comparison, the green line is the constraint found for a constant $\fnl$
          using the same survey assumptions, and the red histograms are the constraints found with a lower $k_{\rm max}$ (see text for details). While the
          individual parameters $\fnl^i$ are poorly constrained as expected,
          their few best linear combinations -- the principal components --
          are well measured; see the next section and text for details. 
}
\label{fig:results}
\end{center}
\end{figure}

We assume a future survey covering one-quarter of the sky (about 10,000
square degrees) out to $z = 1$, and find constraints for a set of 20 $\fnl^i$
uniformly spaced in $\log k$ in the range $10^{-4} \leq k/(\hmpcinv)\leq 1$, with a smoothing scale of
$M_{\rm smooth}=10^{14}\Msun$.
Fig.~\ref{fig:results} shows the resulting unmarginalized (left panel) and marginalized
(right panel) constraints on the parameters $\fnl^i$. 
For both sets of constraints, we first marginalized over the other
cosmological parameters.\footnote{ Using six cosmological parameters along
  with five $b_0(z)$ and 20 $\fnl^i$ led us into some issues with
  floating-point errors and numerical precision. The $31 \times 31$ Fisher
  matrix we obtained was rather ill-conditioned and difficult to invert
  reliably using 64-bit precision; we were eventually forced to move to
  128-bit precision in order to accurately marginalize over the cosmological
  parameters.  }  The $\fnl^i$ have most of their degeneracy
  among themselves; a plot showing the fully unmarginalized constraints on the
  $\fnl^i$ would not look much different than the left panel of Fig.~\ref{fig:results}. Note that, while some of the $\fnl^i$ have support at
$k> k_{\rm max}(z = 1) \approx 0.2\,\hmpcinv$, we only use
information about those (and other) parameters coming from $k<k_{\rm max}$.
The constraints vary considerably as a function of the $k$ at which these
parameters are defined. The best-constrained $\fnl^i$ corresponds to the
$10^{-0.8} < k <10^{-0.6}$ bin, and it has an estimated unmarginalized error
of $\sigma(\fnl^{16}) = 7.3$; for comparison, the worst-constrained $\fnl^i$,
which corresponds to the largest scale (smallest $k$) bin, has an
unmarginalized error well over one billion.

As expected, the marginalized constraints for the best-constrained parameters
are much weaker than the unmarginalized constraints -- even the best-measured
$\fnl^i$ has an estimated marginalized error of $6 \times 10^2$.  In general,
dependence of the constraints on the value of $k$ is determined by two
competing factors: as $k$ increases, there is a larger number of modes, each
with a smaller signal (given by the smaller nongaussian bias $\Delta
b$).  The best-constrained $k$ is also affected by the fact that only
information out to $k=k_{\rm max}=0.1\hmpcinv$ is assumed from the galaxy
survey. In particular, we have checked that if we unrealistically assume
information to be available at all $k$ (instead of at $k< k_{\rm max}$)
without modeling the nonlinearities, the unmarginalized constraints on
$\fnl^{i}$ improve monotonically with increasing $k$. Therefore, the raw
signal-to-noise ratio in $\fnl^{i}$ increases with $k$. To
  further demonstrate the effect of the choice of $k_{\rm max}(z)$, we also
  plotted the errors obtained with the condition $\sigma(\pi/(2k_{\rm max}),
  z) = 0.15$, which yields $k_{\rm max} (z = 0) \approx 0.03$.


\begin{figure}[t] 
\begin{center}
	\subfigure[Unmarginalized errors]{\includegraphics[width= 2.9in]{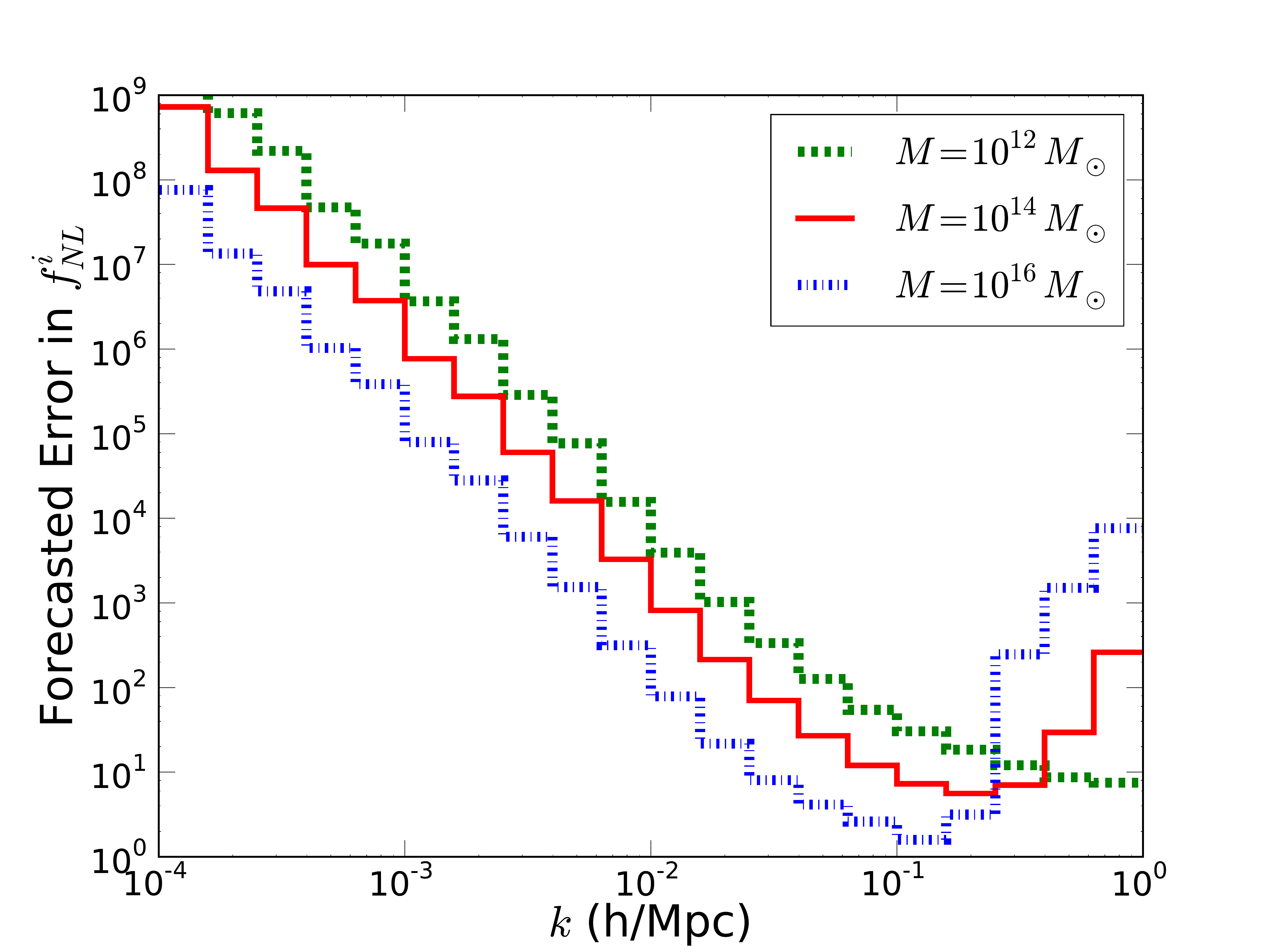}}
	\subfigure[Marginalized errors]{\includegraphics[width= 2.9in]{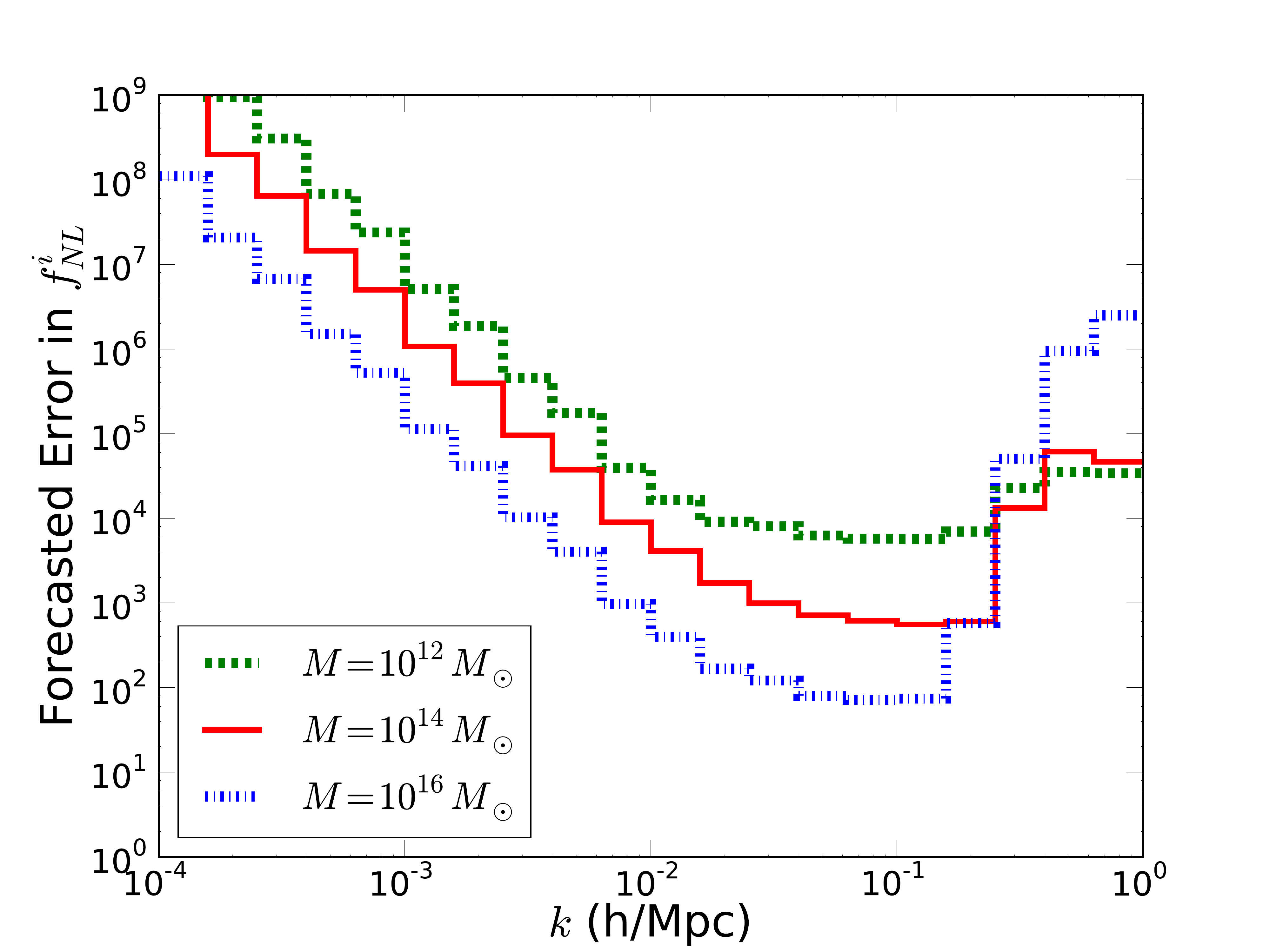}}
        \caption{ Estimated constraints obtained from future surveys with the
          same parameters as the previous figure at different mass smoothing
          scales $M_{\rm smooth}$ (labeled as $M$ in the legend). In other
          words, these are errors for a survey with halos of $M\gtrsim M_{\rm smooth}$.  
}
\label{fig:massdep}
\end{center}
\end{figure}

The smoothing mass scale chosen for this analysis (see
Eq.~(\ref{eq:overdensity})) has a small but noticeable effect on the constraints
yielded. Figure \ref{fig:massdep} shows that, in the case of the
unmarginalized errors, the $k$ at which non-Gaussianity is best constrained
decreases as the smoothing mass scale increases. (The behavior of the
marginalized errors is more complicated due to correlations in errors between
neighboring $\fnl^i$.) Since the mass scale is proportional to
the physical scale (to the third power), this means that best-constrained $k$
decreases with increasing smoothing scale $R$, which is exactly what we should
expect. We remind the reader that while a survey filtered at some scale
$M_{\rm smooth}$ contains objects roughly more massive than this scale, in
practice the near-exponentially falling mass function implies that the number
density is dominated with $M\simeq M_{\rm smooth}$ halos.

\section{Projection and Principal Components}
\label{sec:PC}

\subsection{Constraining other $\fnl(k)$ models}
\label{sec:other_models}

Once the Fisher matrix $F$ has been obtained for the set of parameters
$\fnl^i$, it is quite simple to find the best possible constraints on the
$\fnl^i$ that could be obtained from a future galaxy redshift survey. By
projecting this Fisher matrix into another basis (see Appendix
\ref{app:projection}), it is also possible to find the constraints on any
arbitrary $\fnl(k)$ without calculating a new Fisher matrix from scratch. A
trivial example can be found in Appendix \ref{app:projection}, where we find that
the estimated error on a constant $\fnl$, assuming the same future survey as in the previous section, is
$\sigma(\fnl) = 2.1$. (Note that this forecasted constraint is on a par
with the error expected from Planck, where $\sigma(\fnl) \sim 5$.)

For another, scale-dependent example, consider the simple form of
non-Gaussianity analogous to the conventional parameterization of the power
spectrum
\begin{equation}
\fnl(k) = \fnl^* \left( k \over k_* \right)^{n_{\rm NG}},
\label{eq:runningfnl}
\end{equation}
where $k_*$ is an arbitrary fixed parameter, leaving $\fnl^*$ and $n_{\rm NG}$
as the parameters of interest in this model. ($k_*$ is generally chosen to
minimize degeneracy between $\fnl^*$ and $n_{\rm NG}$ for the observable of
interest. We have set $k_* = 0.165 \hmpcinv$, close to the optimal value in our
case; in CMB analysis, the optimal value is lower, around $0.06 \hmpcinv$.)  The partial
derivatives of our basis of $\fnl^i$ with respect to these parameters are:
\begin{eqnarray}
{\partial \fnl^i \over \partial \fnl^*} &=& \left( k \over k_* \right)^{n_{\rm NG}};
\\[0.2cm]
{\partial \fnl^i \over \partial n_{\rm NG}}  &=& \fnl^* \left( k \over k_* \right)^{n_{\rm NG}} \log \left( k \over k_* \right).
\label{eq:sefusattipartials}
\end{eqnarray}
Starting in a basis of 20 $\fnl^i$ evenly spaced in log $k$, we project down
to a basis of $\fnl^*$ and $n_{\rm NG}$ in order to forecast constraints on the two new
parameters from a survey covering one-quarter of the sky out to $z=1$. We are
using the same limits of integration as in Section \ref{sec:constraints}, along with
the fiducial values $\fnl^* = 50$ and $n_{\rm NG} = 0$.  The forecasted
constraints on these parameters, marginalized over each other, are
$\sigma_{\fnl^*} = 1.7$ and $\sigma_{n_{\rm NG}} = 0.58$.  Despite a
superficial similarity between this model and the model used by Sefusatti et
al.\ in \cite{Sefusatti2009}, the two models are quite different, and our
results cannot be compared. The model used in \cite{Sefusatti2009} is a
function of three arguments, $k_1, k_2$, and $k_3$:
\begin{equation}
\fnl(k_1, k_2, k_3) = \fnl^* \left ({K \over k_*}\right  )^{n_{NG}},
\label{eq:sefusatti_fnl}
\end{equation}
where $K = (k_1 k_2 k_3)^{1/3}$. This leads to a bispectrum of the form found
in Eq.~(\ref{eq:local_bispec}), but with $\fnl(k_1, k_2, k_3)$ in place of $\fnl$,
 whereas our bispectrum is of the less-factorizable form
Eq.~(\ref{eq:fnlk_bispec}).

Another example we consider is the form of non-Gaussianity in which the
running on $\fnl$ itself has running; that is, the case in which $n_{\rm NG}$
is a function of $k$. A simple case of this would be $\fnl$ of the
form\footnote{Analogous parameterization for the power spectrum and its
  motivations are discussed in \cite{kev2}.}
\begin{equation}
\fnl(k) = e^{A k^B}.
\label{eq:runningrunning}
\end{equation}
Projecting the Fisher matrix down from the original basis $\fnl^i$ to the
parameters $A$ and $B$, with fiducial values of $A = \log 50$ and $B = 0$, we
obtain forecasted constraints of $\sigma_A = 1.0$ and $\sigma_B = 0.15$. (In
this case, the survey characteristics and bounds of integration are the same
as in the previous example.)

\subsection{Principal components and relation to local and equilateral models}

\begin{figure}[t]
\begin{center}
\includegraphics[width=6in]{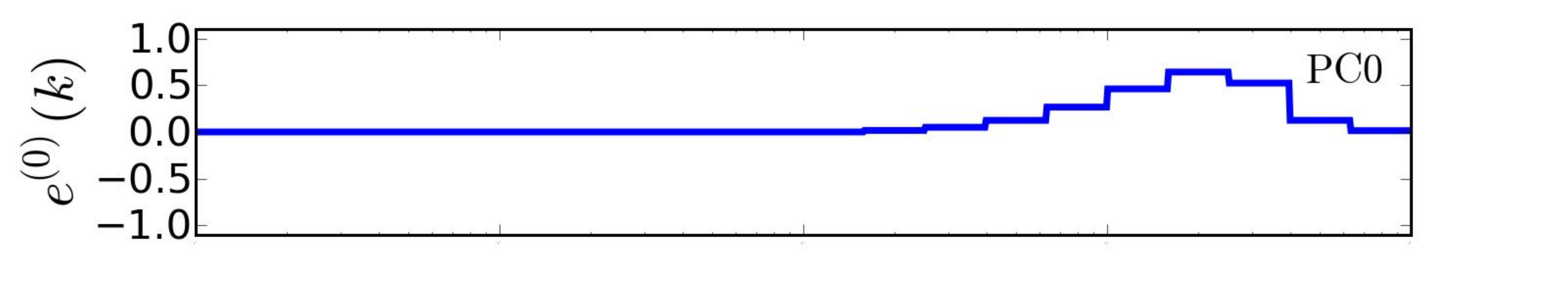} \\[-0.5cm]
\includegraphics[width=6in]{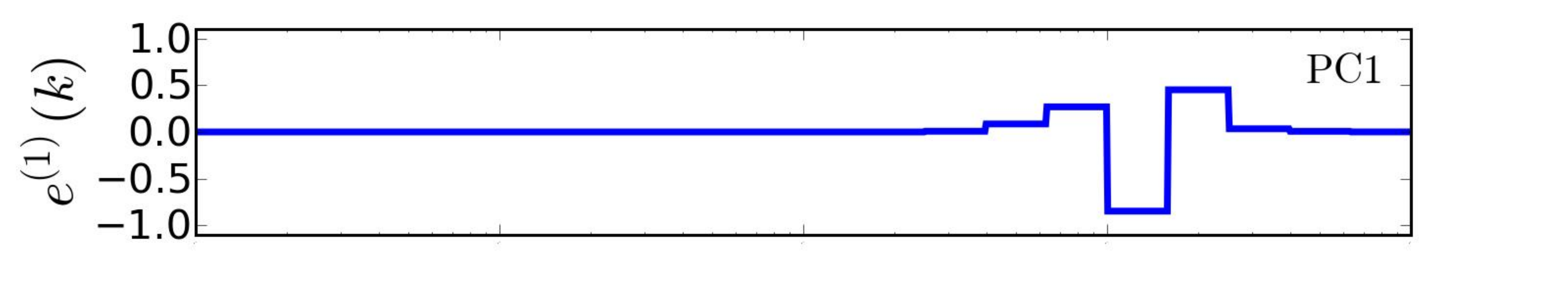} \\[-0.5cm]
\includegraphics[width=6in]{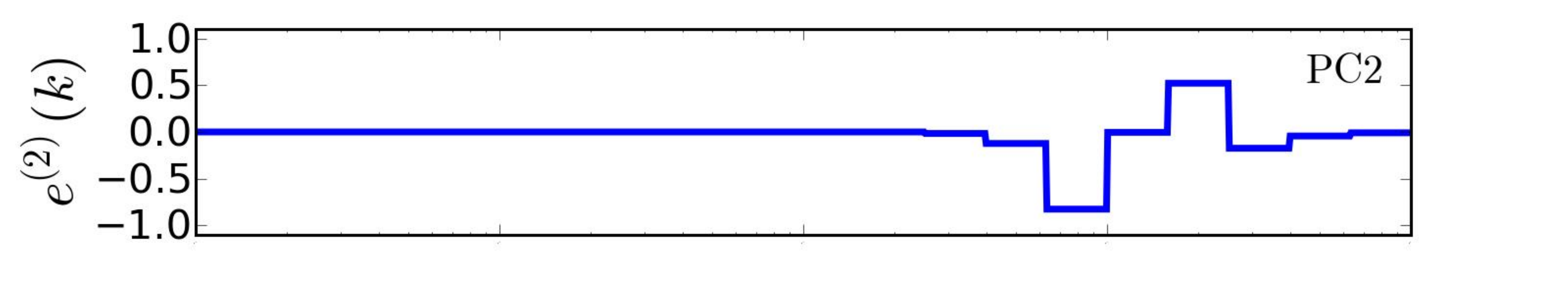} \\[-0.5cm]
\includegraphics[width=6in]{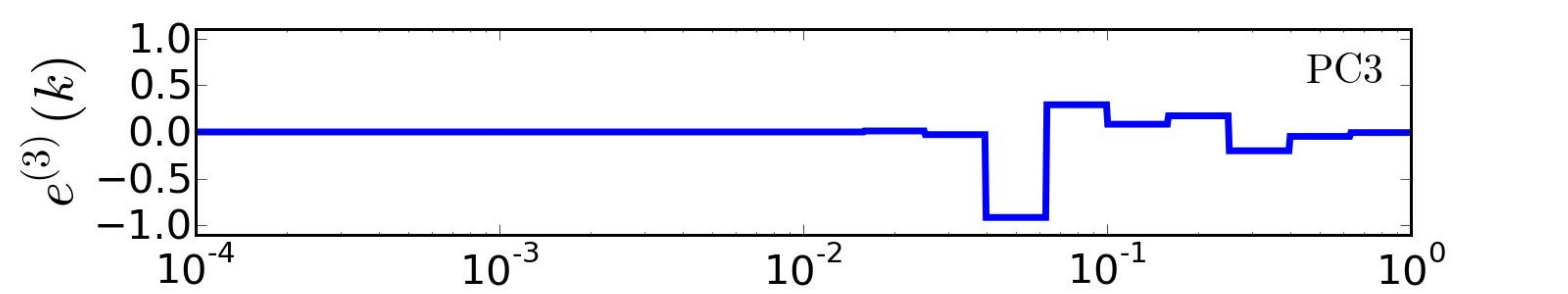}
\end{center}
\caption{The first four principal components of $\fnl(k)$. The PCs,
  $e^{(j)}(k)$, are eigenvectors of the Fisher matrix for the $\fnl^i$, and
  are ordered from the best-measured one ($j=0$) to the worst-measured one
  ($j=19$) for the assumed fiducial survey.  }
\label{fig:unsplinedPC}
\end{figure}

We now represent a general function $\fnl(k)$ in terms of
principal components (PCs). In this approach, the {\it data} determine which
particular modes of $\fnl(k)$ are best or worst measured. The PCs also
constitute a useful form of data compression, so that one can keep only a few of
the best-measured modes to make inferences about the function
$\fnl(k)$. Finally, the PCs will also enable us to measure the degree of
similarity between our scale-dependent ansatz and the local and equilateral
forms of non-Gaussianity.

\begin{figure}[t] 
\begin{center}
\includegraphics[width= 6.2in]{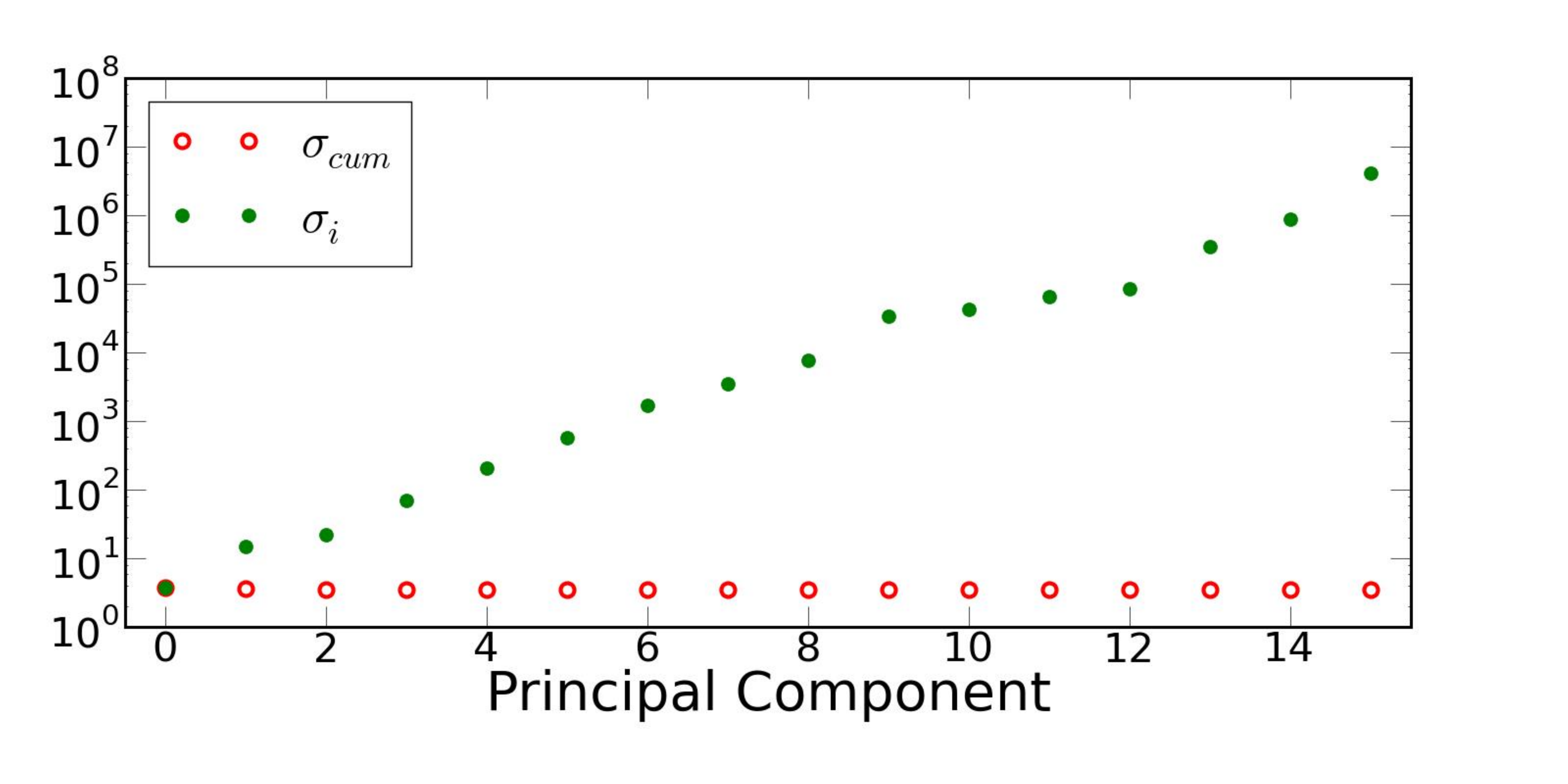}
        \caption{ RMS error on each principal component, along with the
          cumulative error.  }
\label{fig:cumerrs}
\end{center}
\end{figure}

It is rather straightforward to start from the covariance matrix for the
piecewise constant parameters $\fnl^i$ and obtain the PCs of $\fnl(k)$. The PCs are weights in wavenumber with amplitudes that
are uncorrelated by construction, and they are ordered
from the best-measured ($i=0$) to the worst-measured ($i=19$) for the
assumed fiducial survey.  The construction of the PCs is described in Appendix
\ref{app:PC}.  A few of these PCs of $\fnl(k)$ are shown in
Fig.~\ref{fig:unsplinedPC}. For example, the best-measured PC has most of its
weight around $k=10^{-0.4}\,\hmpcinv$, which agrees with sensitivities of
piecewise-constant parameters shown in Fig.~\ref{fig:results}. The sensitivity is not greatest at the largest value of $k$ ($1\,\hmpcinv$) because we assumed cosmological information from $k\leq k_{\rm
  max}=0.1\,\hmpcinv$. We checked that information available at a higher
$k_{\rm max}$ would shift the ``sweet spot" of sensitivity to higher wavenumbers.

The error in the best-measured PC is 4.8; however, the error in the next-best
measured PCs are 18.3 and 27.4, and the accuracy rapidly drops off from
there. Thus, the first three or four PCs should be enough for any conceivable
application. The error in each PC is plotted on a logarithmic scale in figure
\ref{fig:cumerrs}, along with the cumulative error $\sigma_{\rm cum}$, which
is defined as
\begin{equation}
{1 \over \sigma^2_{\rm cum}} = \sum_{i} {1 \over \sigma^2_i}.
\label{eq:cum_error}
\end{equation}

 Each PC $e^{(j)}(k)$ has its own associated bispectrum (see Eq.~(\ref{eq:fnlk_bispec})):
 \begin{equation}
 \label{eq:pc_bispec} 
 B^{(j)}(k_1, k_2, k_3) =
   2[ e^{(j)}(k_1) P(k_2)P(k_3) + e^{(j)}(k_3) P(k_1)P(k_2) + e^{(j)}(k_2) P(k_3)P(k_1) ].
 \end{equation}
(As always, $k_1, k_2,$ and $k_3$ have a triangle relation: $ k_3 = | \vec{k_2} - \vec{k_1} | $.) 
We would like to test the similarity of these bispectra to
those that have already been discussed in the literature. We can do this by
using a distance measure between bispectra, defined by `cosines'
developed in \cite{Babich_shape}.  A cosine near unity implies that the two
bispectra have very similar shapes, and a cosine near zero implies the
opposite. The cosine is defined as
\begin{equation}
\label{cosine} 
\cos(B_1, B_2) = \frac{B_1
    \cdot B_2} { \sqrt{ \left( B_1 \cdot B_1 \right) \left( B_2 \cdot B_2
      \right) }},  
\end{equation}
where the inner product between two bispectra, $B_1 \cdot B_2$, is
\cite{Sefusatti2009}
\begin{equation}
\label{innerprod} 
B_1 \cdot B_2 = \sum_{k_1, k_2, k_3} \frac{ B_1(k_1,
    k_2, k_3) B_2(k_1, k_2, k_3) } { \Delta^2 B(k_1, k_2, k_3) }. 
\end{equation}
The (Gaussian) variance of the bispectrum is
\begin{equation}
\label{gaussvar}
\Delta^2 B(k_1, k_2, k_3)
= \frac{1}{N_T} P(k_1)P(k_2)P(k_3)
\sim \frac{1}{N_T} (k_1 k_2 k_3)^{-3},
\end{equation}
where $N_T$ is the number of distinct triangular configurations of
$k_{1,2,3}$, and $P(k) \sim k^{-3}$ is the primordial curvature perturbation
power spectrum. (The overall constant is irrelevant, since it cancels out in
Eq.~(\ref{cosine}).)

\begin{table}[t]
\centering
\begin{tabular}
{ | c | c | c | }
\hline
 & Local cosine & Equilateral cosine \\
 \hline
$B^{(0)}$ & 0.669 & 0.074 \\  
 \hline
$B^{(1)}$ & 0.040 & 0.000 \\   
 \hline
$B^{(2)}$ & 0.099 & 0.030 \\ 
 \hline
$B^{(3)}$ & 0.189 & 0.037 \\  
 \hline
\end{tabular}
\caption{Cosines of the first four
  principal-component derived bispectra with the local bispectrum and the equilateral
  bispectrum. A cosine near unity implies that the two bispectra have very
  similar shapes, and a cosine near zero implies the opposite. Note that the zeroth PC, which is by far the best
  measured (see Fig.\ \protect\ref{fig:cumerrs}), has a much larger overlap with the local model than with the
  equilateral, as expected. }
\label{unsplinedcosines}
\end{table}

We first compare our bispectra Eq.~(\ref{eq:pc_bispec}) to the local model with
a constant $\fnl$, whose bispectrum is (see Eqs.~(\ref{eq:localNG}) and
(\ref{eq:local_bispec}))
\begin{equation}
\label{squeezedB}
B_{\rm local}(k_1, k_2, k_3) \propto \frac{1}{k_1^3 k_2^3} + \frac{1}{k_1^3 k_3^3} + \frac{1}{k_2^3 k_3^3}.
\end{equation}
Most of the power of $B_{\rm local}$ is in so-called ``squeezed'' triangles,
in which one side is much smaller than the other two (comparable) sides, $k_1
<< k_2 \approx k_3$.

Another form for the bispectrum much discussed in the literature is the
``equilateral" bispectrum
\begin{equation}
\label{equiB}
B_{\rm equi}(k_1, k_2, k_3) = - \frac{2}{(k_2 k_1 k_3)^2} - B_{\rm local}(k_1, k_2, k_3) + \frac{1}{k_1 k_2^2 k_3^3} + \frac{1}{k_3 k_1^2 k_2^3} + \text{permutations}.
\end{equation}
In contrast with $B_{\rm local}$, most of the power of $B_{\rm equi}$ is in
triangles where $k_1 \approx k_2 \approx k_3$; hence the name ``equilateral''.

Table
\ref{unsplinedcosines} lists the cosines of the first few principal-component
derived bispectra with the local bispectrum and the equilateral bispectrum.
The form of Eq.~(\ref{eq:pc_bispec}) suggests that the PC-derived bispectra $B^{(j)}$ will have
more in common with the local bispectrum than the equilateral one. However,
it is initially conceivable that some $e^{(j)}(k)$ might exist which would yield
a bispectrum of the form in Eq.~(\ref{equiB}) when substituted into
Eq.~(\ref{eq:pc_bispec}) -- but in Appendix \ref{sec:ProofAppendix}, we
prove that \textit{no} such function exists. Thus, the only guarantees for the cosines of the $B^{(j)}$
 are that the cosine of $B^{(0)}$ -- the bispectrum corresponding to the best-measured PC -- will
be large with the local model, and that none of the $B^{(j)}$ have a very large
cosine with the equilateral model. We expect the former because our model
looks like the local model; we expect the latter because of the proof in
Appendix \ref{sec:ProofAppendix}. Table \ref{unsplinedcosines} bears out this
expectation. The small cosines with the equilateral form of non-Gaussianity are also unsurprising because equilateral non-Gaussianity is expected to have a strongly suppressed signal in the non-Gaussian halo bias \cite{MV09}.

\section{Conclusions}
\label{sec:concl}

In this paper we have suggested a new phenomenological model of primordial
nongaussianity by generalizing the local model (parametrized with a constant
parameter $\fnl$) to a scale-dependent, non-local class of models. There are
multiple ways to do this, and our choice was to write the Newtonian potential
as 
\begin{equation}
\Phi(x)= \phi_G(x)+\fnl(x)*(\phi_G(x)^2-\langle \phi_G (x)^2 \rangle ),
\end{equation}
where the convolution in real space corresponds to multiplication in
$k$-space, featuring an arbitrary function $\fnl(k)$.  Explicit calculations
show that such a form of the scale dependent $\fnl$ is borne out in inflationary
models \cite{Salopek,Gangui_etal,chris5,chris6,huang}.

We calculated the bispectrum and bias of dark matter halos in this class of
models, following the formalism valid for high peaks
\cite{Grinstein:1986en,MLB1986}.  We then specialized in the
piecewise-constant (in wavenumber) parametrization of $\fnl(k)$ which, for the
case of narrow enough $k$-bins, recovers any arbitrary function. We used
forecasted constraints from an intermediate-future galaxy survey to calculate
errors on individual parameters $\fnl^i$ (see Fig.~\ref{fig:results}) and
briefly studied dependence on the smoothing scale (Fig.~\ref{fig:massdep}).

We further calculated the principal components of $\fnl(k)$, and thus
identified the best-measured configurations (in wavenumber) of this function
(see Fig.~\ref{fig:unsplinedPC}). While the sensitivity increases with
increasing $k$, restricting the survey information to scales where linear
perturbation theory is valid imposes a ``sweet spot'' in sensitivity of
$k\sim 0.1\hmpcinv$.  We then calculated the overlap of the best-measured
principal components with two familiar classes of non-Gaussian models: local
($\fnl={\rm const}$) and equilateral models, using a cosine measure between
the bispectra suggested in \cite{Babich_shape}. We found the expected result:
the best measured component overlaps much more with the local model (which our model generalizes) than with the equilateral one.

One immediate utility of our results is an easy adaptation to specific models of
non-Gaussianity predicted by classes of inflationary models.
If one wants to forecast the
accuracy with which parameters of a specific model of $\fnl(k)$-style non-Gaussianity will be measured, neither
the halo bias nor the Fisher matrix needs to be calculated from scratch. 
Instead, our formalism makes it possible to obtain these
forecasts by performing a simple linear projection to our piecewise-constant model; this
procedure is described in Appendix \ref{app:projection} and illustrated with a
few examples.

In future investigations, it will be interesting to consider specific
inflationary models, projecting down to specific forms for $\fnl(k)$. It will
also be important to test how well the observable effects of scale-dependent
non-Gaussianity, studied here using the theoretical ansatz from
Eq.~(\ref{eq:Grinstein-Wise}), agree with numerical simulations; the first such
investigations, for select specific forms of $\fnl(k)$, are now being done
\cite{Shandera2010}. Finally, it will be interesting to see how one can
optimally select objects in the universe (i.e.\ their mass) to probe
information about scale-dependence of non-Gaussianity. While in
Fig.~\ref{fig:massdep} we showed scaling of the best-determined scale of
$\fnl(k)$ with the smoothing mass scale applied to the density field, a more
complete analysis might use the Halo Occupation Distribution (HOD) approach to
relate the content of dark matter halos to their mass.

\section{Acknowledgements}
We thank  Chris Byrnes and Sarah Shandera for useful discussions, and the anonymous referee for constructive comments.  AB and DH
are supported by DOE OJI grant under contract DE-FG02-95ER40899, NSF under
contract AST-0807564, and NASA under contract NNX09AC89G. KK is supported in
part by the Michigan Center for Theoretical Physics. DH and KK would like to thank the
Aspen Center for Physics where this project germinated, and DH also
acknowledges the generous hospitality of Centro de Ciencias de Benasque ``Pedro
Pascual''.

\appendix

\section{Calculating the error on an arbitrary parametrized $\fnl(k)$}
\label{app:projection}

Projecting the constraints from an old set of parameters $\fnl^i\equiv \fnl(k_i)$
($i=1, 2, \ldots, N$) to new parameters (which we can call $q$; $j=1, 2,
\ldots, M$ for some $M$) is in principle straightforward. The Fisher matrix in
the new parameters, $F^{\rm new}$, is given by
\begin{eqnarray}
F^{\rm new}_{i,j} &= &
\sum_{k,l=1}^N 
\frac{\partial p^k}{\partial q^i}
\frac{\partial p^l}{\partial q^j} F_{kl} 
\label{eq:Fish_transform}
\end{eqnarray}
so that
\begin{equation}
F^{\rm new}  \equiv \mathcal{P}^T F\,\mathcal{P}, 
\label{eq:Fnew}
\end{equation}
where $\mathcal{P}_{i j}=\partial p^i / \partial q^j$ is the derivative
matrix of old parameters with respect to new. 

Let us look at a couple of examples. Projecting to the case 
\begin{equation}
\fnl(k)=\fnl= {\rm  const}
\end{equation}
is particularly easy, since $\mathcal{P}$ is the column vector with
$\mathcal{P}_{i1}=d\fnl^i/d\fnl=1$. Then $F^{\rm new}_{ij}$ is a $1\times 1$
matrix that quantifies information on $\fnl$, given by
\begin{equation}
F_{11}^{\rm new}  = \sum_{k, l}F_{kl}.
\label{eq:Fnew_fnl_const}
\end{equation}
The error on $\fnl$ is of course given simply by $\sigma(\fnl)=1/\sqrt{F_{11}^{\rm new}}$.

Another example is given by the function 
\begin{equation}
\fnl(k)=\left ( {k\over k_0}\right  )^{n_{\rm NG}},
\end{equation}
with two parameters, $k_0$ and $n_{\rm NG}$. Then one can show that (labeling
$k_0\equiv q_1$ and $n_{\rm NG}\equiv q_2$): 
\begin{eqnarray}
\mathcal{P}_{i1} &= & -\frac{n_{\rm NG}}{k_0} \left (\frac{k_i}{k_0}\right )^{n_{\rm NG}};\\[0.2cm]
\mathcal{P}_{i2} &= & \ln\left (\frac{k_i}{k_0}\right )\, \left (\frac{k_i}{k_0}\right )^{n_{\rm NG}}.
\label{eq:M_powerlaw_example}
\end{eqnarray}
Then, using Eq.~(\ref{eq:Fnew}), one can simply obtain the $2\times 2$ Fisher
matrix in $k_0$ and $n_{\rm NG}$.

\section{Principal Components of $\fnl(k)$}
\label{app:PC}

We now show how to decompose the measurement of $\fnl(k)$ in principal
components, which are essentially the eigenmodes of the covariance matrix for
the aforementioned parameters $\fnl(k_i)$. This method has been widely used in
cosmology, including applications to parametrizing and describing dark energy
\cite{Huterer_Starkman,FOMSWG}. It allows us to order the
best-to-worst measured weights in wavenumber of the function $\fnl(k)$.

Let the function $\fnl(k)$ be described in terms of piecewise constant
parameters $\fnl^i\equiv \fnl(k_i)$, where
\begin{equation} \fnl(k) = \sum_{i=1}^N p_i\Theta_i(k).
\end{equation}
Here, $\Theta(k)\equiv \left [H(k-k_i^{\rm lower})- H(k-k_i^{\rm upper})
\right ]$ is the top-hat function of unit height over the $i$th wavenumber
bin, and we assume a total of $N$ bins. $k_i^{\rm lower}$ and $k_i^{\rm
  upper}$ are the wavenumber bin boundaries, and $H$ is the Heaviside step
function. We have effectively expanded the function around the zero value,
though this is not crucial: the left-hand side could be
$\fnl(k)-\fnl^{\rm fid}(k)$, for any fiducial $\fnl^{\rm fid}(k)$, and the
formalism still follows. 

The Fisher matrix $F$ is the inverse covariance matrix in the original
piecewise-constant parameters $p_i$, so that $F^{-1}_{ij}=\langle p_i
p_j\rangle-\langle p_i\rangle\langle p_j\rangle$.
We first diagonalize the Fisher matrix $F$:
\begin{equation}
F = W^T D W,
 \label{eq:F_diag}
\end{equation}
where $D$ is diagonal and $W$ is some orthogonal matrix. The vector of
uncorrelated parameters, ${\bf q}$, is related to the vector of original parameters
${\bf p}$ via
\begin{equation}
{\bf q} = W{\bf p},
\end{equation}
and it is easy to check that the ${\bf q}$ are uncorrelated; that is, $\langle
{\bf q}\, {\bf q}^T\rangle = D^{-1}$. The rows of $W$ are therefore the new
parameters.

Thus, to calculate the principal components:

\begin{enumerate}
\item Obtain the full Fisher matrix for $N$ parameters $p_i$, plus the
  cosmological parameters $\Omega_b h^2, \Omega_{CDM} h^2, H_0, w, \log A_s$, and $n_s$.

\item Marginalize over the cosmological parameters by inverting this larger
  Fisher matrix, taking the $N\times N$ submatrix, then inverting back to get
  the Fisher matrix of the $p_i$; we call {\it this} Fisher matrix $F$

\item Diagonalize $F$ as in Eq.~(\ref{eq:F_diag})

\item The rows of $W$ are the principal components. More precisely,  $q_a = \sum_i
W_{ai}p_i$, and $q_a$ are the PCs. 
\end{enumerate}
Let us now change notation slightly (to agree with the commonly used one,
e.g.~\cite{Huterer_Starkman}), and define the shape of the $a$-th principal
component in $i$-th redshift bin as $\alpha^{(a)}_i$, so that $\alpha^{(a)}_i
\equiv W_{ai}$.  Then we can represent the $a$-th principal component,
$e^{(a)}(k)$, in terms of the original parameters $p_i$ as\footnote{This is
  basically the continuous version of the relation $q_a = \sum_i W_{ai}p_i$.}
\begin{equation}
e^{(a)}(k) = \sum_{i=1}^N \alpha^{(a)}_i p_i \,\Theta_i(k).
\end{equation}
The PCs are obviously uncorrelated, and their eigenvalues $\lambda_a$, so that
\begin{eqnarray}
\langle e^{(a)}e^{(b)}\rangle  
\equiv \sum_{i,j=1}^N\alpha^{(a)}_i\alpha^{(b)}_j \langle p_i p_j\rangle
= \frac{\delta_{ab}}{ \lambda_a}.
\label{eq:orthogonality}
\end{eqnarray}
where, recall, $\lambda_a\equiv D_{aa}$.

Finally, let us calculate the coefficients $c^{(a)}$ in the expansion in
principal components of an arbitrary $\fnl(k)$
\begin{equation}
\fnl(k) = \sum_{a=1}^N c_a e^{(a)}(k).
\label{eq:Fk_ca}
\end{equation}
Let coefficients $\fnl^i$ describe $\fnl(k)$ in our original basis, so that 
$\fnl(k)={\rm const}\equiv \sum_i \fnl^i p_i\Theta_i(k)$, with $\fnl^i$
being left arbitrary for now. Then, taking the expectation value of the
product with $e^{(b)}$, we get
\begin{eqnarray}
\langle \fnl(k) e^{(b)}\rangle \equiv {c_b \over \lambda_b}&=& 
\left \langle \left (\sum_{i=1}^N \fnl^i\, p_i\right )\times \left (\sum_{j=1}^N \alpha_j^{(a)}p_j\right
)\right\rangle\\[0.2cm]
&=& \sum_{i,j=1}^N \fnl^i\, \alpha_j^{(a)} (F^{-1})_{ij},
\end{eqnarray} 
so that
\begin{equation}
c_a =  \lambda_a\,\sum_{i,j=1}^N \fnl^i\, \alpha_j^{(a)}\, (F^{-1})_{ij}.
\end{equation}
For example, in the simplest case of constant $\fnl(k)$, where $\fnl^i={\rm
  const}\equiv \fnl$, the coefficients of the principal components in the
expansion of $\fnl(k)$ are
\begin{equation}
c_a = \lambda_a\,\fnl\,\sum_{ij} \alpha_j^{(a)}\,
(F^{-1})_{ij} \qquad ({\rm for}\,\,\fnl(k)\equiv\fnl={\rm const}). 
\end{equation}

\section{Generalized local ansatz does not recover the equilateral case}
\label{sec:ProofAppendix}

Here, we prove that our ansatz cannot perfectly mimic the equilateral bispectrum for any
choice of $\fnl (k)$.  The generalized local form of the
bispectrum that we considered in this paper is
\eqn{fnlBispec}{
B_{\rm gener}(k_1, k_2, k_3) = 2[\fnl(k_1) P(k_2)P(k_3) + \text{permutations} ] \propto \frac{\fnl(k_1)}{k_2^3 k_3^3}  + \rm perm.
}
The equilateral bispectrum is
\eqn{equiB_redux}{
B_{\rm equi}(k_1, k_2, k_3) \propto 
\left[ \frac{1}{k_1 k_2^2 k_3^3} + \text{perm.} \right] - \frac{2}{(k_2 k_1 k_3)^2} - \left[ \frac{1}{k_2^3 k_3^3} + \text{perm.} \right].
}
The claim is that there is no $\fnl(k)$ such that $B_{\rm gener} = B_{\rm equi}$ for all $k_1, k_2, k_3$.
To show this, we define a new function $h(k) \equiv \fnl(k) +
1$. If there is some $\fnl(k)$ such that $B_{\rm gener} = B_{\rm equi}$, then we have:
\[
\frac{h(k_1)}{k_2^3 k_3^3}  + \text{perm}. \propto 
\left[ \frac{1}{k_1 k_2^2 k_3^3} + \text{perm.} \right] - \frac{2}{(k_2 k_1 k_3)^2}.
\]
We can go from a proportionality to an equality by defining a new function $g(k)$ that is simply $h(k)$ with the appropriate constant out in front.  Next, multiply both sides by $k_1^3 k_2^3 k_3^3$ to get
\eqn{proofeqn}{
k_1^3 g(k_1) + k_2^3 g(k_2) + k_3^3 g(k_3) = 
\left[ k_1 k_2^2 + k_2 k_3^2 + \text{perm.} \right] - 2 k_1 k_2 k_3.
}
Each term on the left-hand side is dependent on only one of $k_1, k_2$, or
$k_3$. However, every term on the right-hand side depends on at least two
different $k$; thus, there is no $g(k)$ that can satisfy this relation.

Alternatively, consider the case where $k_1 = k_2 = k_3 = k$. Then \eqref{proofeqn} becomes
\[
3k^3 g(k) = 4 k^3
\]
which means that
\[
g(k) = 4/3.
\]
This answer is wholly independent of $k$, so this value of $g(k)$ must be true
for all $k$. But this solution for $g(k)$ is clearly incorrect in the general
case where $k_1 \neq k_2 \neq k_3$; therefore, no such $g(k)$ can exist.

While this proves that there is no $f_{\rm NL}(k)$ that yields an exact
equality between our ansatz and the equilateral bispectrum, the question of an
approximate equality remains. Such solutions for $f_{\rm NL}(k)$ certainly
exist for narrow ranges of $k$. For example, $f_{\rm NL}(k) = \delta(k -
k^*)$, where $\delta(k)$ is the Dirac delta function, yields a bispectrum that
is larger for \textit{exactly one} equilateral triangle -- the triangle where
$k_{1, 2, 3} = k_*$ -- than it is for any squeezed triangle. However, no
$f_{\rm NL}(k)$ exists that yields a bispectrum which favors equilateral
triangles over squeezed triangles for all $k$. It is straightforward but
tedious to prove this fact, and the details of the proof are beyond the scope
of this paper.

\bibliographystyle{JHEP}

\bibliography{fnl}

\end{document}